\documentclass{webofc}

\usepackage[varg]{txfonts}   
\usepackage{hyperref}
\usepackage{url}
\hypersetup{colorlinks=true,citecolor=blue,urlcolor=blue,linkcolor=blue}
\begin{document}
\title{Mechanisms of $D \bar{D}$ pair production in $e^{+} e^{-} \to e^{+} e^{-} D \bar{D}$ and $AA \to AA D \bar{D}$ reactions and production of exotic charmonia}
%
%

\author{\firstname{Antoni} \lastname{Szczurek}\inst{1, 2}\fnsep\thanks{\email{Antoni.Szczurek@ifj.edu.pl}} \and
        \firstname{Izabela} \lastname{Babiarz}\inst{1}
\fnsep\thanks{\email{Izabela.Babiarz@ifj.edu.pl}} \and
        \firstname{Piotr} \lastname{Lebiedowicz}\inst{1}\fnsep\thanks{\email{Piotr.Lebiedowicz@ifj.edu.pl}} \and
        \firstname{Wolfgang} \lastname{Sch\"afer}\inst{1}\fnsep\thanks{\email{Wolfgang.Schafer@ifj.edu.pl}}
}

\institute{Institute of Nuclear Physics Polish Academy of Sciences, \\
Radzikowskiego 152, PL-31342 Krak\'ow, Poland
\and
           Institute of Physics, Faculty of Exact and Technical Sciences, University of Rzesz\'ow,\\
Pigonia 1, PL-35310 Rzesz\'ow, Poland
          }

\abstract{We discuss mechanisms of $D \bar{D}$ pair production in $e^+ e^-$ collisions.
We consider the $D^0 {\bar D}^0$-continuum mechanism 
with the $t/u$-channel $D^*$ vector-meson exchanges,
where the $D^* D \gamma$ coupling constants are derived from the $D^* \to D \gamma$ decays.
For the $D^+ D^-$ channel, we consider mainly the $D^{\pm}$ exchanges. 
We find a relatively large continuum contribution for the $D^0 {\bar D}^0$ channel 
and a significantly smaller one for $D^+ D^-$, 
concluding that the broad bump at $M_{D^0 {\bar D}^0} = 3.8$~GeV
observed by the Belle and BaBar Collaborations has a continuum origin 
rather than corresponding to the broad $\chi_{c0}(3860)$ resonance.
We also investigate the resonant production of $D \bar{D}$ pairs,
including $\chi_{c0}(3860)$ and $\chi_{c2}(3930)$.
The $\chi_{c2}(3930)$ meson is a good candidate 
for the first excited $P$-wave charmonium $\chi_{c2}(2P)$ state.
This resonance can decay into both $D \bar{D}$ channels; however,
its branching fractions are not well known at present.
Comparing our model result with $\Gamma(\chi_{c2}(2P) \to \gamma \gamma) = 0.544$ keV
(corresponding to the Buchm\"uller-Tye potential) to the BaBar data,
we find $B(\chi_{c2}(3930) \to D \bar{D}) = 0.58 \pm 0.13$.
We present results for differential distributions in several variables for Belle~II kinematics.
Finally, we shortly mention predictions for the ultraperipheral heavy-ion reaction $Pb Pb \to Pb Pb D \bar{D}$
for low invariant mass ($M_{D \bar{D}} < 4.3$ GeV) at the LHC.}

\maketitle
%

\section{Introduction}
\label{sec:Introdution}

The properties of ground state $P$-wave charmonia 
$\chi_{cJ}$ ($J = 0, 1, 2$) are relatively well known. 
In contrast, the excited $P$-wave charmonia are poorly understood.
The production of axial $\chi_{c1}(2P)$ state
was discussed for $pp$ \cite{Cisek:2022uqx} 
and $e^+ e^-$ \cite{Babiarz:2023ebe} collisions.
A number of  states were considered as candidates for the first 
excited $P$-wave charmonia $\chi_{c0,2}(2P)$.
A candidate for the $\chi_{c2}(2P)$ state ($\chi_{c2}(3930)$)
was observed by the Belle Collaboration \cite{Belle:2005rte} 
and later confirmed by the BaBar Collaboration \cite{BaBar:2010jfn} 
in the two-photon fusion process 
$e^+ e^- \to e^+ e^- (\gamma \gamma \to D \bar{D})$
as a resonance 
in the $D^{+} D^{-}$ and $D^0 \bar{D}^0$ systems.
In the $D^0 {\bar D}^0$ channel,
one also observes a broad enhancement at $M \sim 3.8$~GeV.
A more detailed discussion of the problem was presented
in our recent paper \cite{Babiarz:2025sld}
(where all details can be found).
Here we show selected results from reference~\cite{Babiarz:2025sld}. 

In reference~\cite{Babiarz:2025sld}, we have considered two different mechanisms
for the $e^{+} e^{-} \to e^{+} e^{-} D \bar{D}$ reactions. 
The first mechanism is the so-called continuum contribution 
described by $t$- and $u$-channel meson exchanges and contact terms.
For the $D^0 {\bar D}^0$ final state
we have considered the vector-meson $D^{*}(2007)^{0}$ $t/u$-channel 
exchanges as the underlying continuum process
while in the $D^+ D^-$ case
we have considered the pseudoscalar $D^{\pm}$ $t/u$-channel exchanges 
and a contact term.
The second production mechanism proceeds through an intermediate 
$s$-channel resonance states.
We consider the $\chi_{c0}(3860)$ and $\chi_{c2}(3930)$ resonances, 
assuming that they correspond to $\chi_{c0}(2P)$ and $\chi_{c2}(2P)$, 
respectively.
Our aim in reference~\cite{Babiarz:2025sld} was to describe the Belle and BaBar data 
simultaneously, for both the neutral and charged channels separately.

In reference~\cite{Lebiedowicz:2026qff}, the production of $D \bar{D}$ pairs
in ultraperipheral heavy-ion reaction ${\rm Pb} {\rm Pb} \to {\rm Pb} {\rm Pb} D \bar{D}$ was considered,
limiting to low invariant mass of $D \bar{D}$ and for energies available at the LHC.
There, the total cross section and several differential distributions 
for experimental cuts corresponding 
to the ALICE, ATLAS, CMS, and LHCb experiments are presented. 

\section{Exclusive production of $D \bar{D}$ pairs in $e^+ e^-$ collisions}
\label{sec:Formalism}

\begin{figure}[h]
\centering
\includegraphics[width=0.3\textwidth]{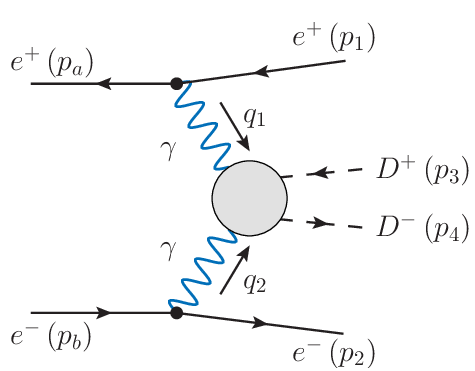}
\quad
\includegraphics[width=0.64\textwidth]{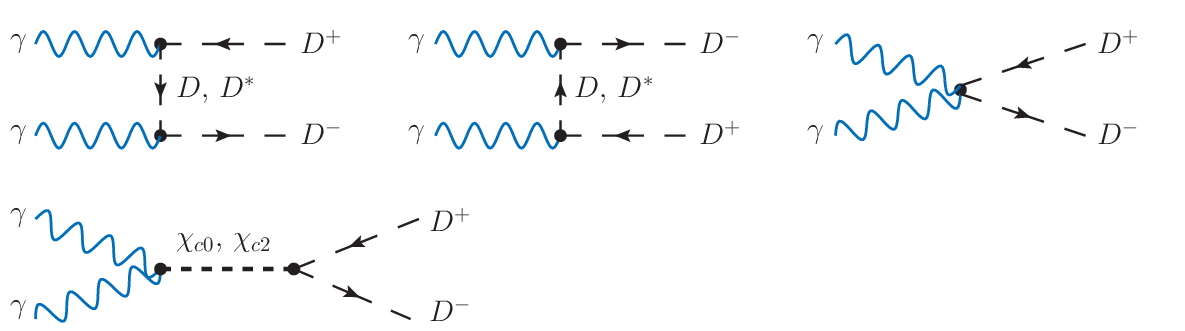}
\caption{Diagrams representing $D^{+}D^{-}$ production in $e^{+}e^{-}$ collisions.
For the process $\gamma \gamma \to D^{+} D^{-}$ we consider 
the $t$- and $u$-channel $D^{\pm}$ and $D^{*}(2010)^{\pm}$ exchanges,
the contact term,
and the exchange of $\chi_{c0,2}$ mesons.
The diagrams for the $D^0 {\bar D}^0$ production are similar, 
but in the case of non-resonant production
we consider the $D^{*}(2007)^{0}$ exchanges.}
\label{fig:diagram1}
\end{figure}

We have performed  calculations for the exclusive
$2 \to 4$ reaction;
see the diagrams shown in figure~\ref{fig:diagram1}.
The fully differential cross section can be written as
\begin{eqnarray}
d \sigma = \frac{1}{2 \sqrt{s (s - 4m_{e}^{2})}} 
\overline{|{\cal M}_{2 \to 4}|^{2}} (2 \pi)^{4}
\delta^{(4)}(p_{1} + p_{2} + p_{3} + p_{4} - p_{a} - p_{b}) 
\prod_{i=1}^{4}
\frac{d^{3}p_{i}}{(2\pi)^{3}2p_{i}^{0}}\,.
\label{xs_2to4}
\end{eqnarray}
Here, $p_{a}, p_{b}, p_{1}, \ldots, p_{4}$ are the momenta of incoming and outgoing particles,
$s = (p_{a} + p_{b})^{2}$,
$m_{e}$ is the electron mass, and
$\overline{|{\cal M}_{2 \to 4}|^{2}}$ is the amplitude squared 
averaged over initial and summed over final lepton polarization states.
To calculate the total cross section one has to perform integration in
an eight dimensional phase space.

We considered the photon-photon fusion production mechanism 
relevant for small-angle scattering of leptons. 
The corresponding $2 \to 4$ amplitude has the form
\begin{eqnarray}
{\cal M}_{\lambda_a \lambda_b \to \lambda_1 \lambda_2 D \bar D} = 
i j^{(1)}_{\alpha}(\lambda_1, \lambda_a) \, 
\frac{- i g^{\alpha \mu}}{q_1^2} \, 
{\cal M}_{\mu \nu}(\gamma^* \gamma^* \to D \bar D) \, 
\frac{-i g^{\nu \beta}}{q_2^2} \, 
i j^{(2)}_\beta(\lambda_2,\lambda_b) \,,
\end{eqnarray}
where $\lambda_{a}, \lambda_{b}, \lambda_{1}, \lambda_{2} \in \{1/2, -1/2 \}$
are the lepton polarizations, and
$q_{1} = p_{a} - p_{1}$, 
$q_{2} = p_{b} - p_{2}$.
${\cal M}_{\mu \nu}(\gamma^* \gamma^* \to D \bar D)$
describes both resonance production and continuum mechanisms.
Below, we also use the Mandelstam variables
$t_{1} = q_{1}^{2}$, $t_{2} = q_{2}^{2}$, and
$p_{34} = q_{1} + q_{2} = p_{3} + p_{4}$. 
Furthermore, for the lepton currents, we have
\begin{align}
j^{(1)}_{\alpha}(\lambda_1, \lambda_a)  &= \bar{v}(p_{1}, \lambda_{1})
\gamma_{\alpha}
v(p_{a}, \lambda_{a}) 
\approx (p_{1} + p_{a})_{\alpha} \,\delta_{\lambda_1 \lambda_a}
\,, \nonumber \\
j^{(2)}_{\beta}(\lambda_2, \lambda_b) &= 
\bar{u}(p_{2}, \lambda_{2}) 
\gamma_\beta 
u(p_{b}, \lambda_{b})
\approx (p_{2} + p_{b})_{\beta} \,\delta_{\lambda_2 \lambda_b}
\,.
\end{align}
%

\subsection{Resonant production: $\gamma^* \gamma^* \to \chi_{c2} \to D \bar{D}$}
\label{sec:Formalism_A}

The amplitude for the 
$\gamma^*(q_1) \gamma^*(q_2) \to \chi_{c2} \to D(p_3) \bar{D}(p_4)$ 
subprocess is given by \cite{Babiarz:2025sld}
\begin{eqnarray}
{\cal M}_{\mu \nu}(\gamma^{*} \gamma^{*} \to \chi_{c2} \to D \bar{D}) 
=
i\Gamma^{(\gamma^* \gamma^* \to \chi_{c2})}_{\mu \nu \rho \sigma}(q_{1},q_{2}) \;
i\Delta^{(\chi_{c2})\,\rho \sigma, \alpha \beta}(p_{34})\;
i\Gamma^{(\chi_{c2} \to D \bar{D})}_{\alpha \beta}(p_{3},p_{4})\,.
\label{3.1}
\end{eqnarray}
Above, we have introduced the
propagator for a narrow tensor-meson resonance in
the relativistic Breit-Wigner form \cite{Babiarz:2025sld}
\begin{eqnarray}
i\Delta_{\mu \nu, \kappa \lambda}^{(\chi_{c2})}(p_{34}) =
\frac{i}{p_{34}^{2}-M_{\chi_{c2}}^2+i M_{\chi_{c2}} \Gamma_{\chi_{c2}}}
\left[ 
\frac{1}{2} 
( \hat{g}_{\mu \kappa} \hat{g}_{\nu \lambda}  + \hat{g}_{\mu \lambda} \hat{g}_{\nu \kappa} )
-\frac{1}{3} 
\hat{g}_{\mu \nu} \hat{g}_{\kappa \lambda}
\right] \,,
\label{3.2}
\end{eqnarray}
where $\hat{g}_{\mu \nu} = -g_{\mu \nu} + p_{34 \mu} p_{34 \nu} / p_{34}^2$.
The total decay width of the $\chi_{c2}(3930)$ resonance and its mass
($M_{\chi_{c2}} = 3922.5 \pm 1.0~{\rm MeV}$,
$\Gamma_{\chi_{c2}} = 35.2 \pm 2.2~{\rm MeV}$)
are taken from the PDG \cite{PDG}.

In \cite{Babiarz:2025sld} we limited ourselves only to helicity-2 component, 
The corresponding vertex can be written as
\begin{equation}
\Gamma_{\mu \nu \kappa \lambda}^{(\gamma^* \gamma^* \to \chi_{c2})}(q_{1},q_{2})=
e^2 
\frac{1}{2} \Big( R_{\mu \kappa}  R_{\nu \lambda}  
                + R_{\nu \kappa}  R_{\mu \lambda}
                - R_{\mu \nu}  R_{\kappa \lambda}  \Big) F_{TT,2}(Q_1^2,Q_2^2)
 \,,
\label{2.1}
\end{equation}
where $e^{2} = 4 \pi \alpha_{\rm em}$,
the photon virtualities
$Q_{1,2}^{2} \equiv -q_{1,2}^{2} \geqslant 0$,
and 
%
\begin{equation}
R_{\mu \nu} \equiv R_{\mu \nu}(q_1,q_2)
= -g_{\mu \nu} + \frac{1}{X}
\Big( (q_{1} \cdot q_{2}) (q_{1 \mu} q_{2 \nu} + q_{1 \nu} q_{2 \mu})
- q_{1}^{2} q_{2 \mu} q_{2 \nu} - q_{2}^{2} q_{1 \mu} q_{1 \nu} \Big)\,,
\label{2.2}
\end{equation}
$X = (q_{1} \cdot q_{2})^{2} - q_{1}^{2} q_{2}^{2}$. 
The transition form factor was parametrized as
\begin{eqnarray}
F_{TT,2}(Q_1^2,Q_2^2) = F_{TT,2}(0,0) 
\frac{\Lambda^2}{Q_1^2 + Q_2^2 + \Lambda^2} \,.
\label{2.7}
\end{eqnarray}
We assume the $\chi_{c2}(3930)$ state to be $\chi_{c2}(2P)$ state
containing pure $c \bar{c}$ component.
One can calculate $F_{TT,2}(0,0)$ in the NRQCD limit 
[see (4.2) of \cite{Babiarz:2024sqw}]
\begin{equation}
F_{TT,2}(0,0) = 8 e_f^2 \sqrt{\frac{3 N_c}{\pi M_{\chi_{c2}}^{3}}} |R_{2P}'(0)|
\label{2.12}
\end{equation}
with $e_f = 2/3$ and $N_{c} = 3$.
The first derivative of the radial wave function at the origin [$R_{2P}'(0)$]
is estimated for several $c \bar{c}$ interaction potential models, 
see table~I of \cite{Babiarz:2025sld}.
The transition form factor
$F_{TT,2}(0,0)$ is related to the two-photon decay 
width of the $\chi_{c2}$ state,
\begin{eqnarray}
\Gamma(\chi_{c2}(2P) \to \gamma \gamma) =
(4 \pi \alpha_{\rm em})^2 \frac{|F_{TT,2}(0,0)|^2}{80 \pi M_{\chi_{c2}}}\,.
\end{eqnarray}
Using $|R_{2P}'(0)| = (0.228 - 0.405)$~GeV$^{5/2}$ \cite{Babiarz:2025sld},
we find $\Gamma(\chi_{c2}(2P) \to \gamma \gamma) \equiv
\Gamma_{\gamma \gamma}(\lambda = \pm 2) =
 (0.27 - 0.84)$~keV,
assuming the dominance of the helicity-2 term.
Specifically, the value $\Gamma_{\gamma \gamma} = 0.544$ keV
corresponds to to the Buchm\"uller-Tye potential.

The decay width of $\chi_{c2} \to D^{+} D^{-}$ is given by
\begin{eqnarray}
\Gamma(\chi_{c2} \to D^{+} D^{-}) = 
\frac{M_{\chi_{c2}}}{480 \pi}\,
|g_{\chi_{c2} D\bar{D}}|^{2} 
\left( \frac{M_{\chi_{c2}}}{M_{0}} \right)^{2}
\left( 1-\frac{4 m_{D}^{2}}{M_{\chi_{c2}}^{2}} \right)^{5/2}\,.
\label{3.6}
\end{eqnarray}
Here $M_{0} \equiv 1$~GeV.
We take 
$\Gamma(\chi_{c2} \to D \bar{D}) = 14 - 26$~MeV,
and using (\ref{3.6}) and assuming isospin invariance in the decay
[$\Gamma(\chi_{c2} \to D^{+} D^{-}) = 
\Gamma(\chi_{c2} \to D^{0} \bar{D}^{0}) = 
\frac{1}{2}\Gamma(\chi_{c2} \to D \bar{D})$],
we obtain
$|g_{\chi_{c2} D\bar{D}}| = 8.34 - 11.37$.
With $\Gamma_{\chi_{c2}} = 35.2$~MeV
we get the branching fraction in the range
$B(\chi_{c2} \to D \bar{D}) = 0.40 - 0.74$;
see \cite{Babiarz:2025sld}.

\subsection{Resonant production: $\gamma^* \gamma^* \to \chi_{c0} \to D \bar D$}
\label{sec:Formalism_chic0}

We include also resonant production of scalar quarkonium: 
$\gamma^* \gamma^* \to \chi_{c0} \to D \bar D$.
We consider two scenarios. In the first scenario:
\begin{eqnarray}
M_{\chi_{c0}} = 3862^{+50}_{-35}~{\rm MeV}\,, \qquad
\Gamma_{\chi_{c0}} = 200^{+180}_{-110}~{\rm MeV}\,;
\label{4.0_pdg}
\end{eqnarray}
see \cite{PDG}.
In the second scenario:
\begin{eqnarray}
M_{\chi_{c0}} = 3862~{\rm MeV}\,, \qquad
\Gamma_{\chi_{c0}} = 50~{\rm MeV}\,.
\label{4.0_fit}
\end{eqnarray}
The scalar-meson propagator is parametrized as
\begin{eqnarray}
i\Delta^{(\chi_{c0})}(p_{34})=
\frac{i}{p_{34}^{2}-M_{\chi_{c0}}^2+i M_{\chi_{c0}} \Gamma_{\chi_{c0}}}
\label{4.1a}
\end{eqnarray}
with a constant decay width.

The $\gamma^* \gamma^* \to \chi_{c0}$ vertex,
limiting only to the transverse component,
can be written as
\begin{equation}
\Gamma_{\mu \nu}^{(\gamma^* \gamma^* \to \chi_{c0})}(q_{1},q_{2}) =
-e^2 
R_{\mu \nu}
F_{TT}(Q_1^2,Q_2^2)\,.
\label{4.2}
\end{equation}
Here, $R_{\mu \nu}$ is defined by (\ref{2.2}) and the form factor $F_{TT}$ is parametrized as
\begin{eqnarray}
F_{TT}(Q_1^2,Q_2^2) = F_{TT}(0,0) 
\frac{\Lambda^2}{Q_1^2 + Q_2^2 + \Lambda^2} \,, \quad \Lambda = M_{\chi_{c0}}\,.
\label{2.3a}
\end{eqnarray}
Assuming that the $\chi_{c0}(3860)$ state is the $c\bar{c}$ state one can calculate
$F_{TT}(0,0)$ in the NRQCD limit as 
[see (4.11) of \cite{Babiarz:2020jkh}]
\begin{equation}
F_{TT}(0,0) = e_f^2 \sqrt{N_c} \frac{12}{\sqrt{\pi}}
\frac{R_{2P}'(0)}{M_{\chi_{c0}}^{3/2}}\,;
\label{4.4}
\end{equation}
see table~II of \cite{Babiarz:2025sld}.
For the $\chi_{c0} D \bar{D}$ vertex we have ($M_{0} \equiv 1$~GeV)
\begin{eqnarray}
i\Gamma^{(\chi_{c0} \to D \bar{D})}(p_{3},p_{4})=
i g_{\chi_{c0} D\bar{D}} M_{0} \,
F^{(\chi_{c0})}(p_{34}^{2})\,,
\label{4.6}
\end{eqnarray}
where the coupling constant $g_{\chi_{c0} D\bar{D}}$ is related to the partial
decay width of the $\chi_{c0}$ meson
\begin{eqnarray}
\Gamma{(\chi_{c0} \to D \bar{D})}=
\frac{M_{0}^{2}}{16 \pi M_{\chi_{c0}}} |g_{\chi_{c0} D\bar{D}}|^{2} 
\left( 1-\frac{4 m_{D}^{2}}{M_{\chi_{c0}}^{2}} \right)^{1/2}\,.
\label{4.7}
\end{eqnarray}
Taking $M_{\chi_{c0}} = 3862$~MeV and
$\Gamma(\chi_{c0} \to D \bar{D}) = 16.4$~MeV we obtain
$|g_{\chi_{c0} D\bar{D}}| = 2.52$;
see also the discussion after (2.33) of \cite{Babiarz:2025sld}.

\subsection{Continuum production of neutral $D$ mesons: $\gamma^{*} \gamma^{*} \to D^{0} \bar{D}^{0}$}

The $\gamma \gamma \to D^{0} \bar{D}^{0}$ amplitude
with the $D^{*0} \equiv D^{*}(2007)^{0}$ 
$t/u$-channel exchanges can be expressed as \cite{Babiarz:2025sld}
\begin{align}
{\cal M}_{\mu \nu}(\gamma^{*} \gamma^{*} \to D^0 \bar D^0)
&=
i\Gamma^{(D^{*0}D^0\gamma)}_{\kappa_{1} \mu}(\hat{p}_{t},q_{1})\,
i\tilde{\Delta}^{(D^{*0})\,\kappa_{1} \kappa_{2}}(s_{34},\hat{p}_{t}^{2})\,
i\Gamma^{(D^{*0}D^0\gamma)}_{\kappa_{2} \nu}(-\hat{p}_{t},q_{2}) 
\nonumber \\
& \quad + 
i\Gamma^{(D^{*0}D^0\gamma)}_{\kappa_{1} \mu}(-\hat{p}_{u},q_{1})\,
i\tilde{\Delta}^{(D^{*0})\,\kappa_{1} \kappa_{2}}(s_{34},\hat{p}_{u}^{2})\,
i\Gamma^{(D^{*0}D^0\gamma)}_{\kappa_{2} \nu}(\hat{p}_{u},q_{2}) \,,
\label{amplitude_D0D0bar_continuum}
\end{align}
where 
$\hat{p}_{t} = p_{a} - p_{1} - p_{3}$,
$\hat{p}_{u} = p_{4} - p_{a} + p_{1}$, and $s_{34} = p_{34}^{2}$.
We use in (\ref{amplitude_D0D0bar_continuum})
$\tilde{\Delta}^{(D^{*0})\, \kappa_{1} \kappa_{2}}(s_{34},\hat{p}^{2})$
the propagators
for the reggeized vector meson $D^{*}(2007)^{0}$.
Here, we take into account the vector-meson reggeization effect analogously to (3.20)--(3.22) of \cite{Lebiedowicz:2019jru}.

The $D^{*0}D^{0}\gamma$ vertex, including a form factor, 
is taken in the form:
\begin{eqnarray}
i\Gamma_{\mu \nu}^{(D^{*0}D^0\gamma)}(\hat{p},q) 
= -i e \frac{g_{D^{*0}D^0\gamma}}{m_{D^{*0}}}\, 
\varepsilon_{\mu \nu \rho \sigma} \hat{p}^{\rho} q^{\sigma}
F^{(D^{*0}D^{0}\gamma)}(\hat{p}^{2},q^{2})\,.
\label{DstarDgam_vertex}
\end{eqnarray}
We use 
\begin{eqnarray}
F^{(D^{*0}D^0\gamma)}(\hat{p}^{2},q^{2}) = 
F^{(D^{*0})}(\hat{p}^{2}) F^{(\gamma)}(q^{2}) 
\label{DstarDgam_ff}
\end{eqnarray}
with $F^{(D^{*0}D\gamma)}(m_{D^{*0}}^{2},0) = 1$, and
\begin{eqnarray}
F^{(D^{*0})}(\hat{p}^{2}) &=& \exp \left( \frac{\hat{p}^{2}-m_{D^{*0}}^{2}}{\Lambda_{D^{*0}}^{2}} \right)\,,
\label{DstarDgam_ff_exp}
\\
F^{(\gamma)}(q^{2}) &=& \frac{\Lambda_{\gamma}^{2}}{\Lambda_{\gamma}^{2}-q^{2}}\,,
\qquad \Lambda_{\gamma} = 1\;{\rm GeV}\,.
\label{DstarDgam_ff_aux}
\end{eqnarray}
Here $\hat{p}^{2} < 0$ and $q^{2} < 0$.
We take $\Lambda_{D^{*0}} = 3.3 - 3.5\;{\rm GeV}$
estimated from a comparison of the model results
to the BaBar data (see figure~\ref{fig:M34_BaBar_eff} below).

\subsection{Continuum production of charged $D$ mesons: $\gamma^{*} \gamma^{*} \to D^{+} D^{-}$}
\label{sec:Formalism_C}

The amplitude for the subprocess 
$\gamma \gamma \to D^{+} D^{-}$ has the form:
\begin{align}
{\cal M}_{\mu \nu}(\gamma^{*} \gamma^{*} \to D^{+} D^{-})
&= -i e^{2}
\left[ 
(q_{1} - 2 p_{3})_{\mu}(q_{1} - p_{3} + p_{4})_{\nu} \frac{1}{\hat{p}_{t}^{2} - m_{D}^{2}} 
\right. \nonumber \\
&\quad + \left.  (q_{1} - 2 p_{4})_{\mu}(q_{1} + p_{3} - p_{4})_{\nu} \frac{1}{\hat{p}_{u}^{2} - m_{D}^{2}} 
- 2 g_{\mu \nu} \right]
\nonumber \\
& \quad \times 
F^{(\gamma)}(q_{1}^{2}) F^{(\gamma)}(q_{2}^{2})
F(\hat{p}_{t}^{2},\hat{p}_{u}^{2},p_{34}^{2})\,.
\label{amplitude_DpDm_continuum}
\end{align}
Above $F^{(\gamma)}(q_{1,2}^{2})$ are given by
(\ref{DstarDgam_ff_aux}) and
the off-shell meson dependences can be taken into account by 
multiplication of the amplitude by a common form factor
$F(\hat{p}_{t}^{2},\hat{p}_{u}^{2},p_{34}^{2})$
[see (2.12) and (2.13) of \cite{Klusek-Gawenda:2017lgt}
but with the replacements $m_{p} \to m_{D^{+}}$ and $\Lambda_{p} \to \Lambda_{D}$].
The parameter $\Lambda_{D}$ should be fitted to the experimental data.
In the analysis, we performed calculations for 
$F(\hat{p}_{t}^{2},\hat{p}_{u}^{2},p_{34}^{2}) = 1$.

By a similar way, the $D^{+} D^{-}$-production mechanism 
via the $D^{*}(2010)^{\pm}$ exchanges was considered;
see \cite{Babiarz:2025sld} for more details.

\section{Selected results}
\label{sec:Results}

\begin{figure}[h]
\centering
\includegraphics[width=0.325\textwidth]{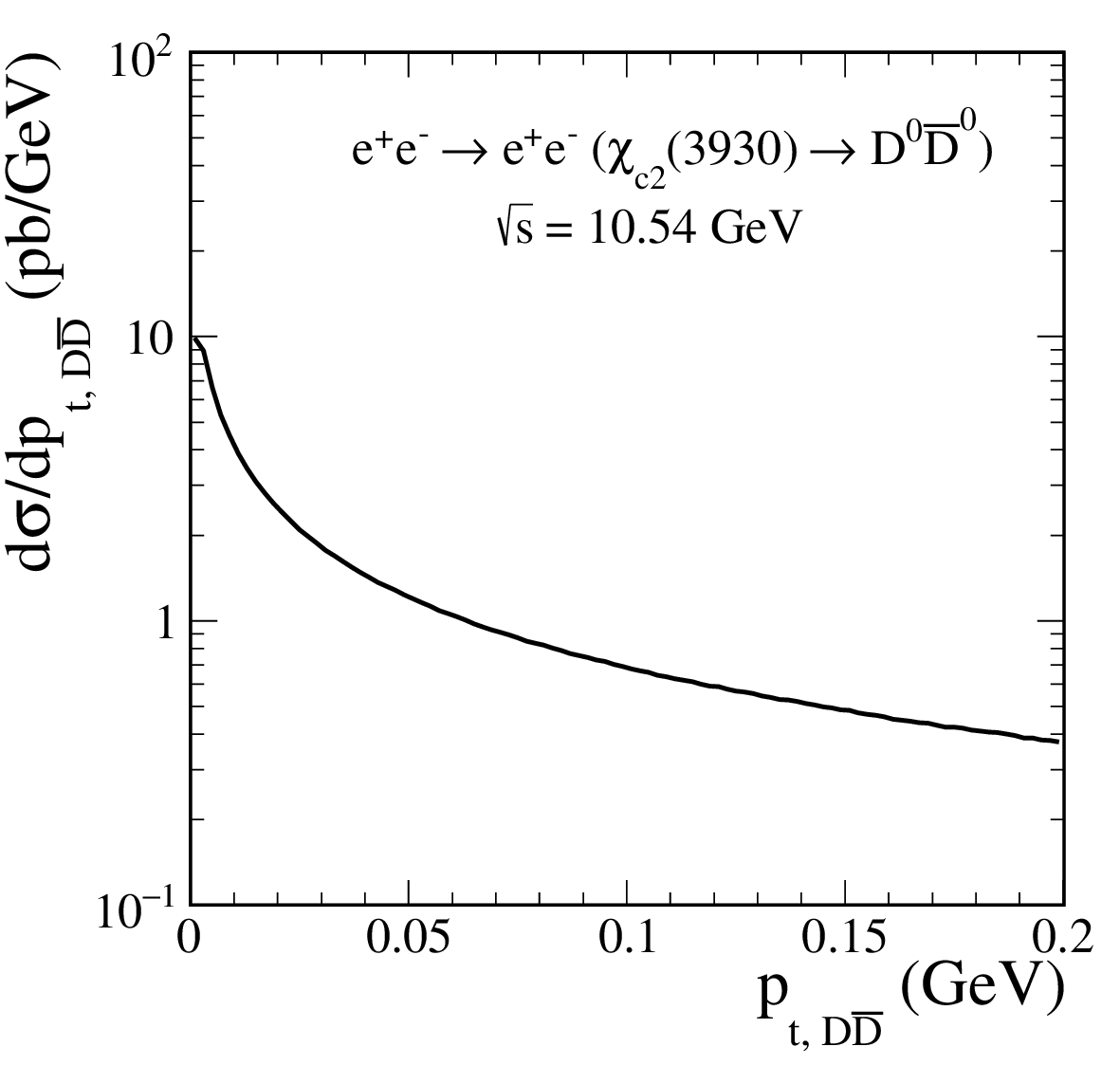}
\includegraphics[width=0.325\textwidth]{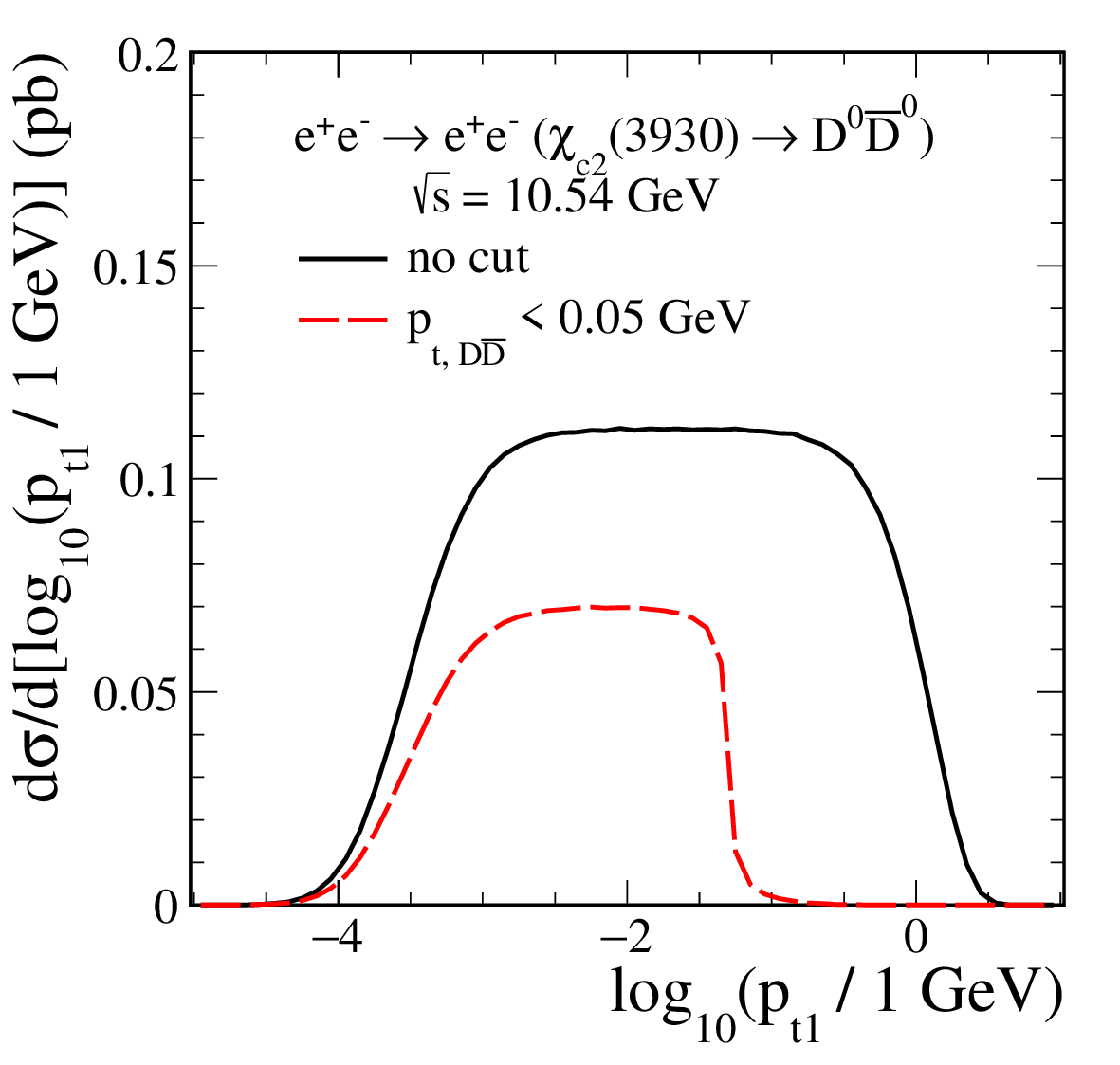}
\includegraphics[width=0.325\textwidth]{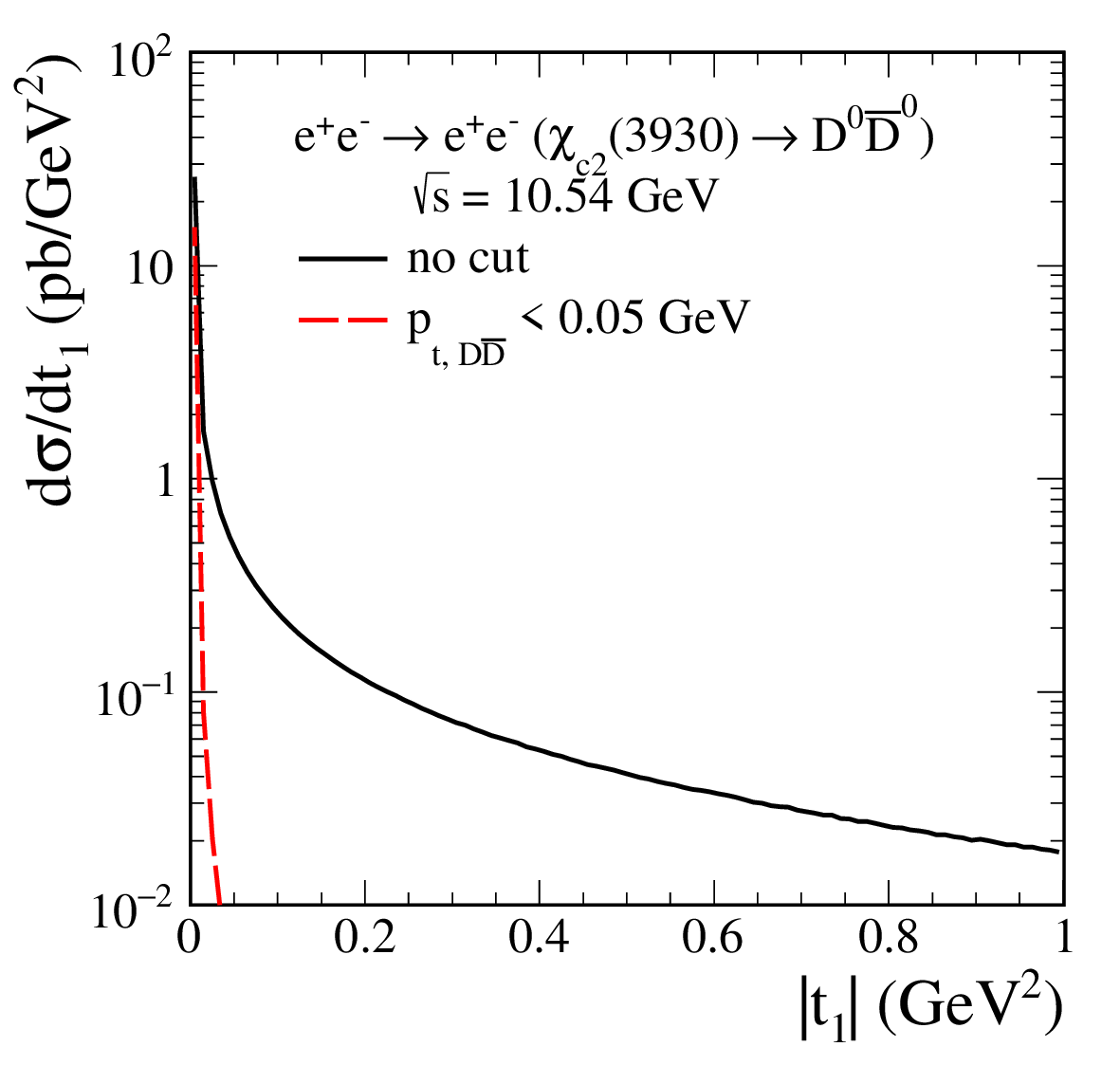}
\caption{Distributions in $p_{t, D\bar{D}}$,
$\log_{10}(p_{t1}/ 1\,\rm{GeV})$ and
$|t_{1}|$ 
for the $e^+ e^- \to e^+ e^- (\chi_{c2}(3930) \to D^0 {\bar D}^0)$ reaction
for $\sqrt{s} = 10.54$~GeV.
We show the results without and with the cut on $p_{t, {\rm sum}} < 0.05$~GeV.
The calculations were done for the Buchm\"uller-Tye potential 
and with
$g_{\chi_{c2} D \bar{D}} = 10.1$.
(From figure~2 of \cite{Babiarz:2025sld}).
}
\label{fig:dsig_dxi}
\end{figure}

Here we wish to show some differential distributions
for the $e^+ e^- \to e^+ e^- (\chi_{c2}(3930) \to D^0 {\bar D}^0)$ reaction.
Figure~\ref{fig:dsig_dxi} shows the distributions in
$p_{t, D\bar{D}}$ the transverse momentum of the $D^0 {\bar D}^0$ pair,
in $\log_{10}(p_{t1}/ 1\,\rm{GeV})$,
one of the integration variables, and
in $|t_{1}|$ the four-momentum transfer squared.
In the calculation
we used $g_{\chi_{c2} D \bar{D}} = 10.1$
corresponding to the Buchm\"uller-Tye (BT) potential
(see table~III of \cite{Babiarz:2025sld}).
We show results without imposing any restrictions on kinematic variables
and the results with the extra condition
$p_{t, D\bar{D}} < 0.05$~GeV.
The distribution in $p_{t, D\bar{D}}$
peaks in the region of low $p_{t, D\bar{D}}$.
The $t_{1}$ and $t_{2}$ distributions are strongly peaked 
at very small $|t_{1}|$ and $|t_{2}|$.
This is caused by the factors $1/t_{1,2}$ occurring in the photon propagators.
One can clearly see that imposing 
the condition $p_{t, D\bar{D}} < 0.05$~GeV
significantly reduces the cross-section
and limits the momentum transfers to the electrons,
$|t_{1,2}| < 0.01$~GeV$^{2}$.

\begin{figure}[h]
\centering
(a)\includegraphics[width=0.45\textwidth]{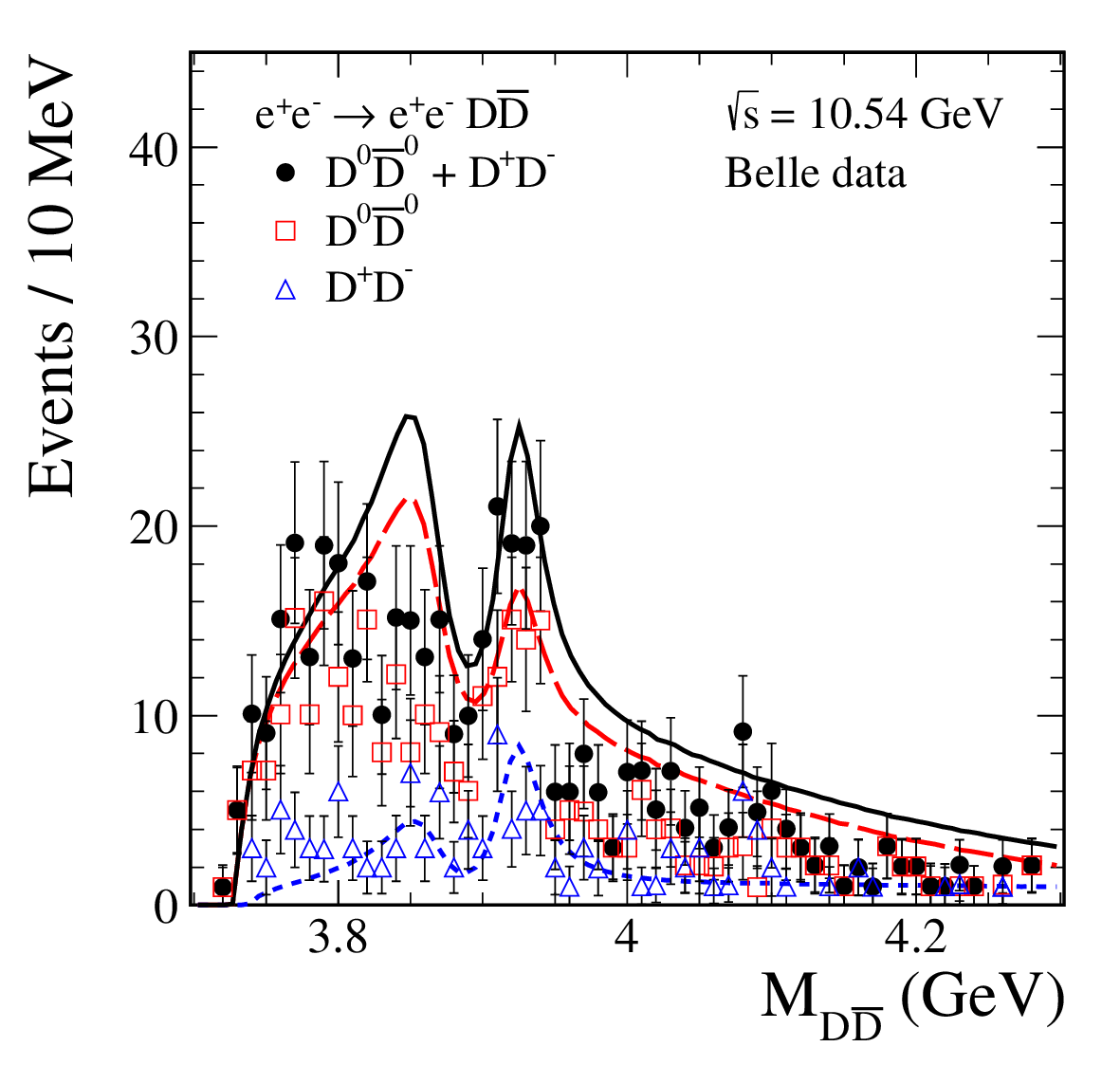}
(b)\includegraphics[width=0.45\textwidth]{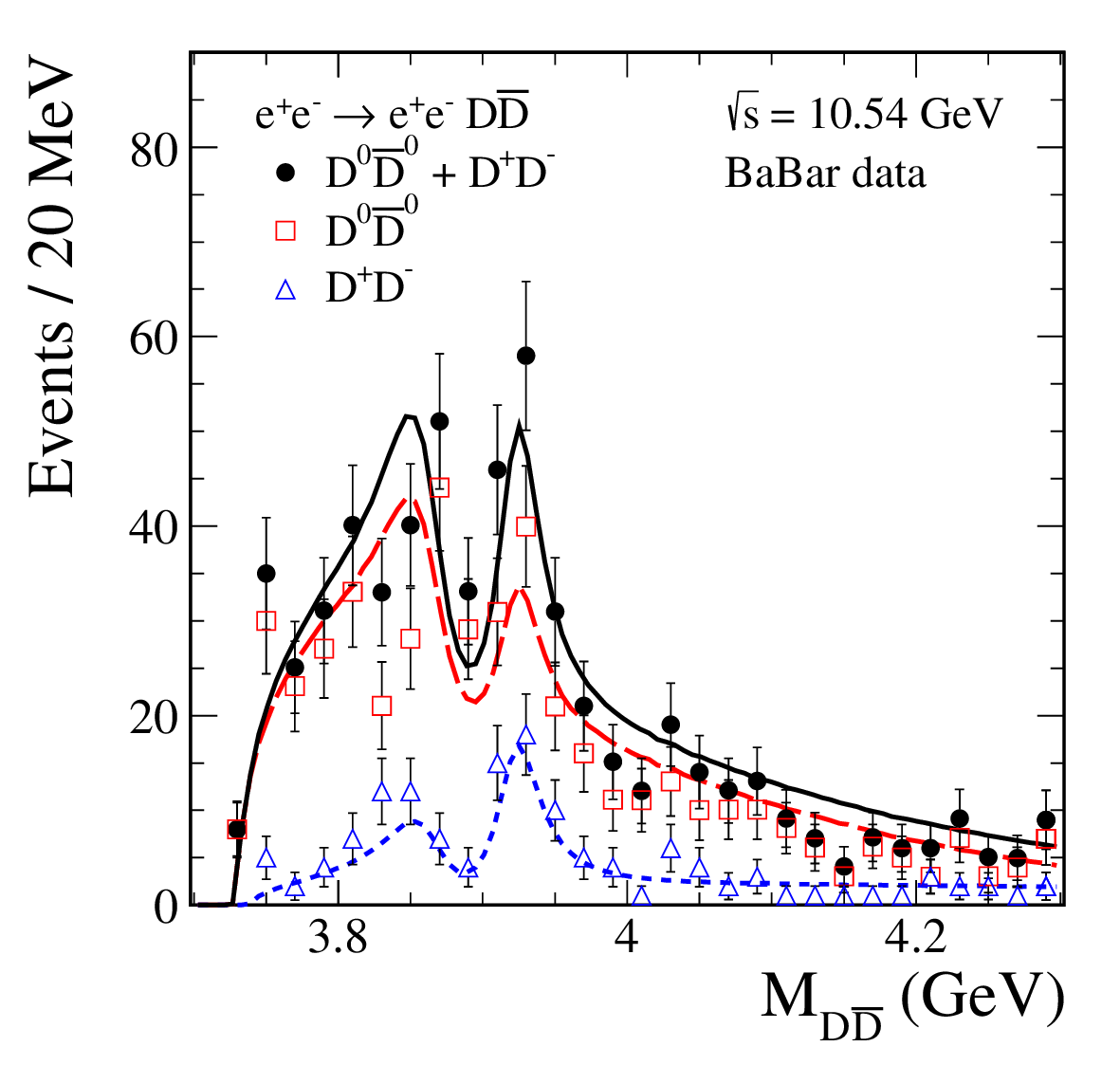}
(c)\includegraphics[width=0.45\textwidth]{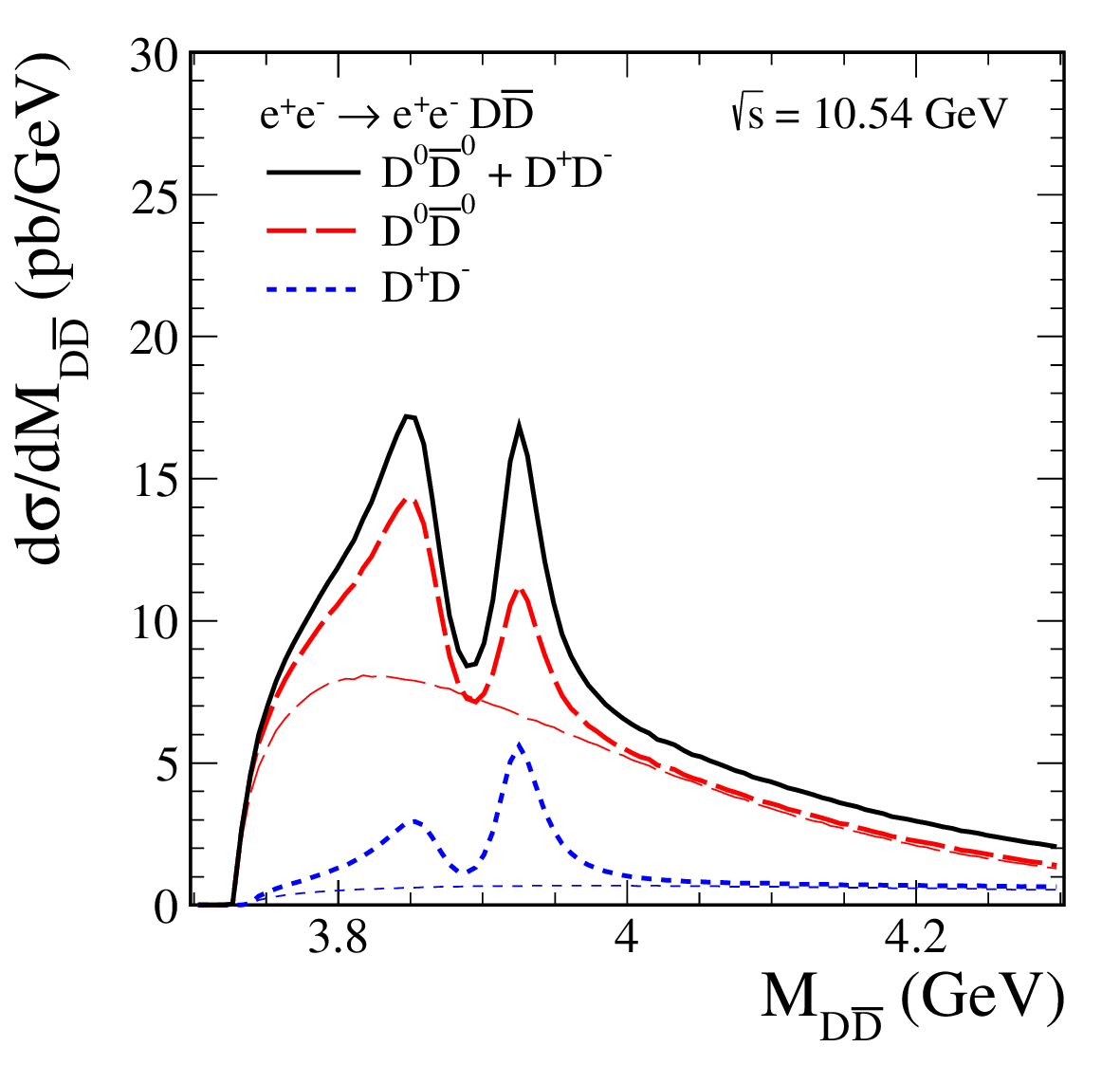}
(d)\includegraphics[width=0.45\textwidth]{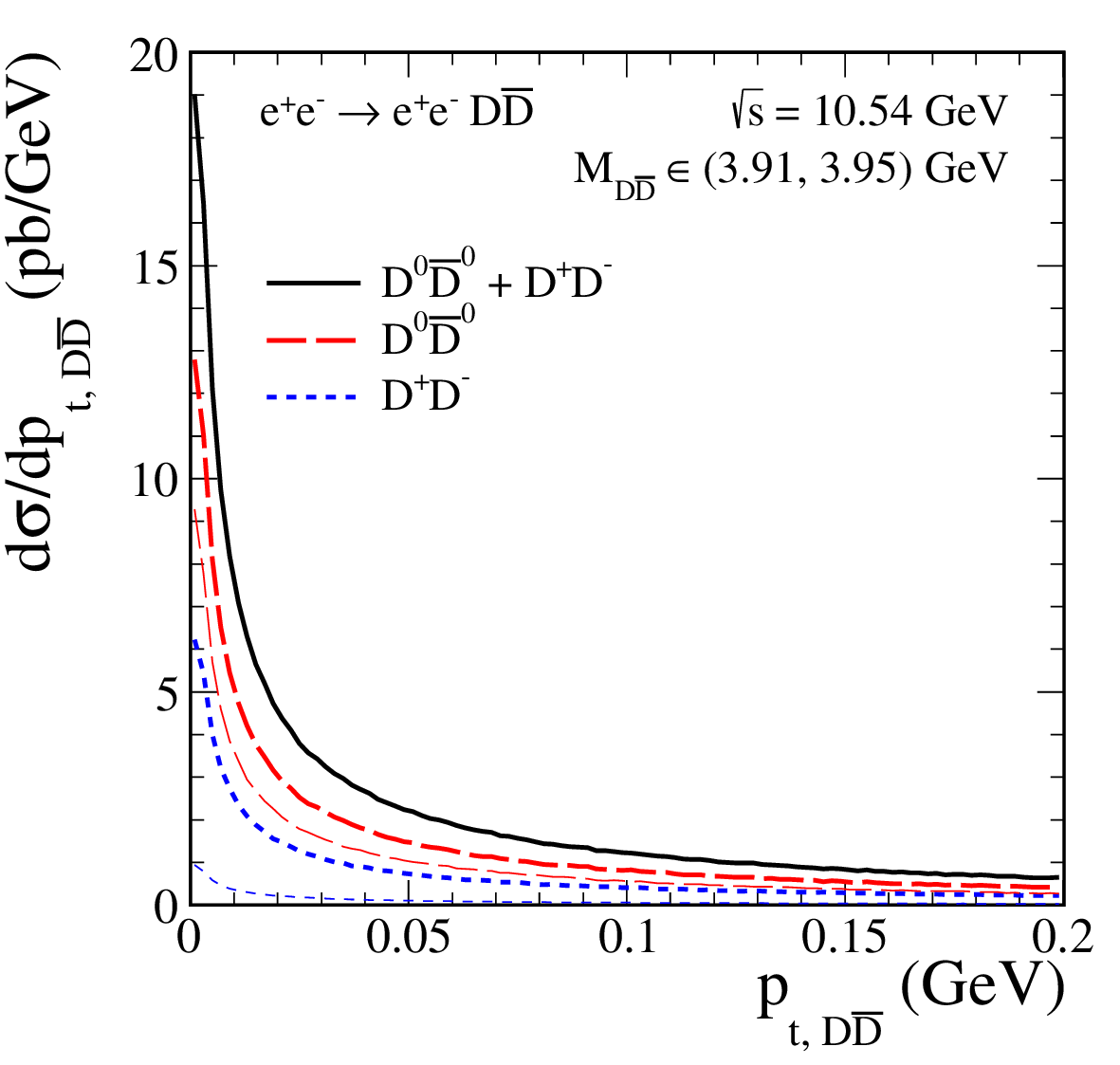}
\caption{
The invariant mass distribution 
for the $e^{+} e^{-} \to e^{+} e^{-} D \bar{D}$ reaction
for the neutral and charged channels,
and the combined $D \bar{D}$ final state.
(a) 
Comparison of theoretical results 
to the Belle data \cite{Belle:2005rte}.
(b) 
Comparison of theoretical results 
to the BaBar data \cite{BaBar:2010jfn}.
(c)
The differential cross sections $d\sigma/dM_{D \bar{D}}$
for the complete model (continuum, $\chi_{c0}(3860)$, and $\chi_{c2}(3930)$ contributions)
and for the continuum contribution alone.
For the $\chi_{c0}(3860)$ contribution we take (\ref{4.0_fit}).
The red long-dashed lines correspond to the $D^0 {\bar D}^0$ channel
and the blue dashed lines correspond to the $D^{+}D^{-}$ channel.
The black solid line represents the sum of all contributions for the two channels.
In panel (d), we show the differential distributions in transverse momentum
of the $D \bar{D}$ pair.
The calculations were done for $\sqrt{s} = 10.54$~GeV
and with the cut on $M_{D \bar{D}} \in (3.91, 3.95)$~GeV.
The meaning of the lines is the same 
as in figure~\ref{fig:M34_BaBar}(c).
(From figure~3 of \cite{Babiarz:2025sld}).
}
\label{fig:M34_BaBar}
\end{figure}

We compare our theoretical results 
for the $e^{+} e^{-} \to e^{+} e^{-} D \bar{D}$ reaction
calculated for $\sqrt{s} = 10.54$~GeV
with the Belle data from figure~2 of \cite{Belle:2005rte}
and
BaBar data from figure~5 of \cite{BaBar:2010jfn}.
In figure~\ref{fig:M34_BaBar}(a)-(c)
we show the invariant mass distributions
for the combined $D \bar{D}$ final state
and for the $D^{0} \bar{D}^{0}$ and $D^{+} D^{-}$ channels separately. 
We can see that the complete model results, 
containing the continuum and resonant contributions, 
reasonably well describes the experimental data.
There is a significant interference effect between 
the $\chi_{c0}(3860)$ resonance and continuum terms.
The $D^{0} \bar{D}^{0}$-continuum contribution
increases from the reaction threshold, 
reaches a maximum at $M_{D^{0} \bar{D}^{0}} \sim 3.8$~GeV, 
and has a large share in the $\chi_{c2}(3930)$ resonance region.
In figure~\ref{fig:M34_BaBar}(d) we show the distribution 
in $p_{t, D\bar{D}}$ 
in the region $M_{D \bar{D}} \in (3.91, 3.95)$~GeV.

\begin{figure}[h]
\centering
\includegraphics[width=0.45\textwidth]{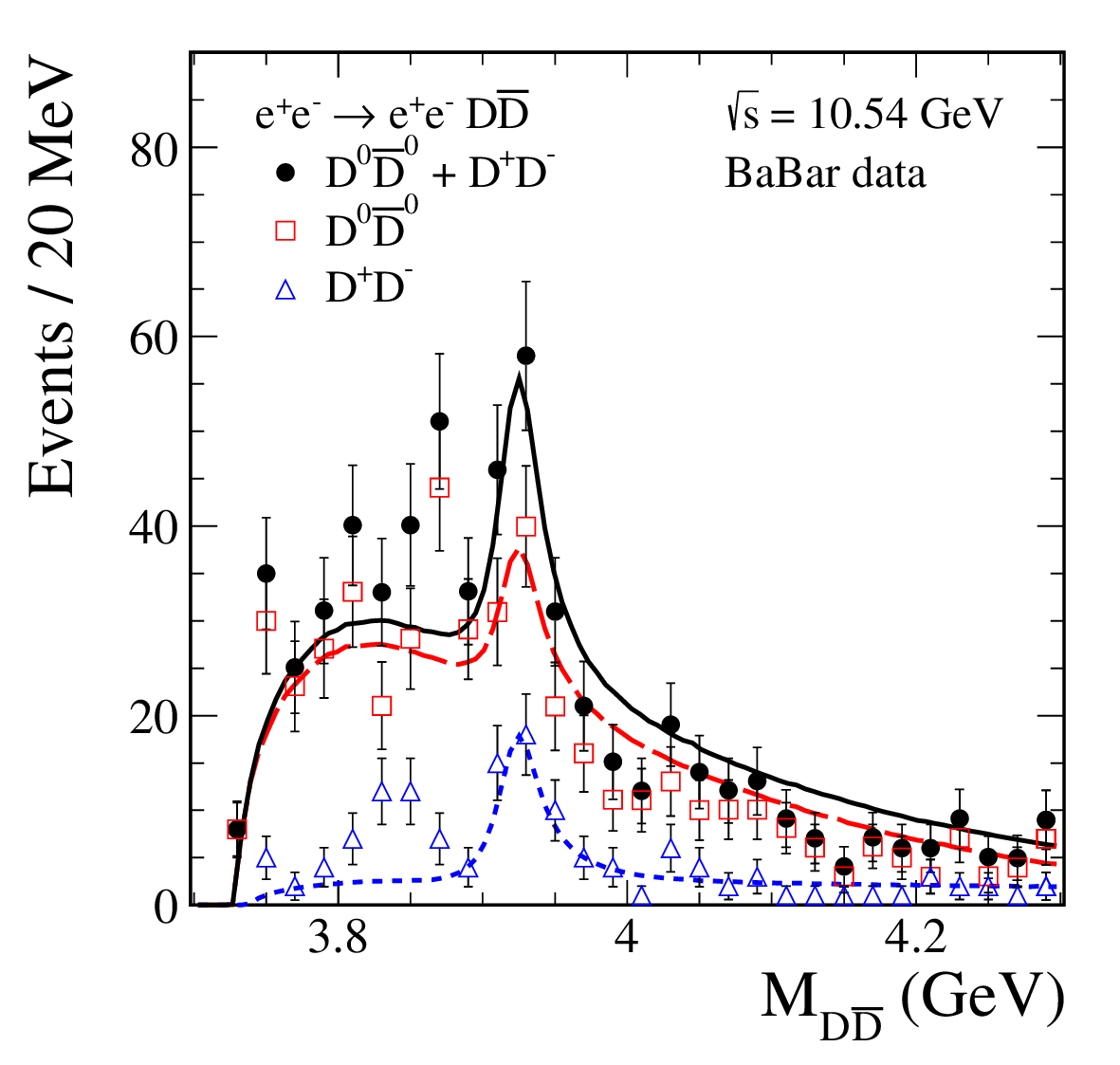}
\includegraphics[width=0.45\textwidth]{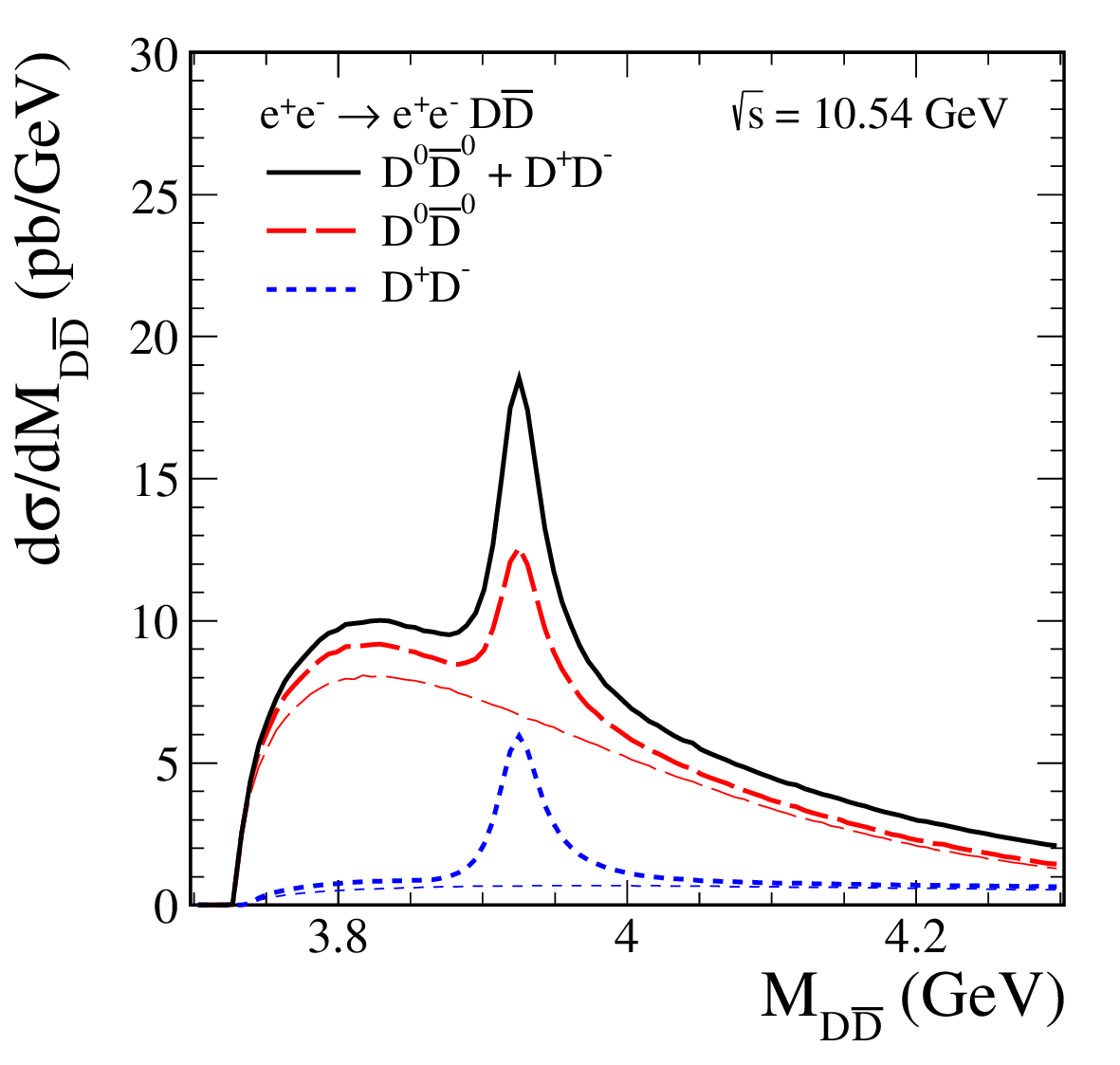}
\caption{
Comparison of theoretical results 
to the BaBar data 
\cite{BaBar:2010jfn}.
The meaning of the lines is as in figure~\ref{fig:M34_BaBar},
but for the $\chi_{c0}(3860)$ contribution
we used the middle values from (\ref{4.0_pdg}).
(From figure~4 of \cite{Babiarz:2025sld}).
}
\label{fig:M34_BaBar_200}
\end{figure}

In figure~\ref{fig:M34_BaBar_200}, we present our results 
obtained for 
$\Gamma_{\chi_{c0}} = 200$~MeV 
(\ref{4.0_pdg}) for the $\chi_{c0}(3860)$ resonance.
In this case, we have that the increase 
in $\Gamma_{\chi_{c0}}$ implies a worse description of low-$M_{D \bar{D}}$ data
(especially for the $D^{+}D^{-}$ BaBar data).
Here, we keep the coupling constants 
the same as in the previous scenario. 
In general, using the larger total width, 
the coupling constant of $\chi_{c0} \to D \bar{D}$ could be 
consistently reestimated. 

From the right panel of figure~\ref{fig:M34_BaBar_200}
we see that the contributions from the $\chi_{c2}(3930)$ resonance
and $D^{0} \bar{D}^{0}$-continuum are dominant.
The contribution from $\chi_{c0}(3860)$ is found to be small
but the interference effect between the $\chi_{c0}(3860)$
and the $D^{0} \bar{D}^{0}$ continuum is rather large.

\begin{figure}[h]
\centering
\sidecaption
\includegraphics[width=0.48\textwidth]{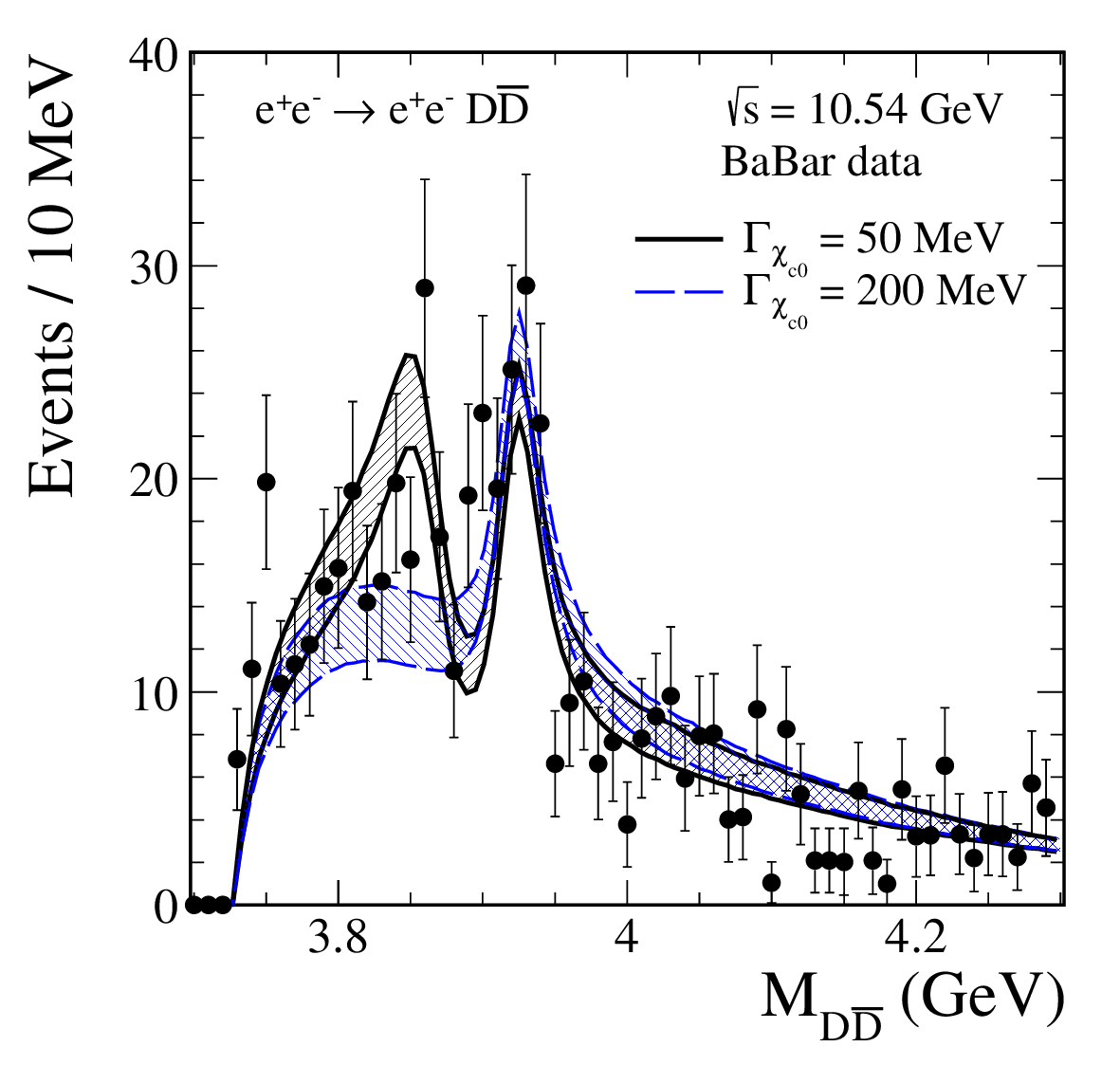}
\caption{
Comparison of our complete model results to the efficiency-corrected 
(combined) BaBar data from \cite{BaBar:2010jfn}.
The black solid lines represent results
obtained for $\Gamma_{\chi_{c0}} = 50$~MeV,
while the blue dashed lines correspond to 
$\Gamma_{\chi_{c0}} = 200$~MeV.
Here, the calculations for the $D^{0} \bar{D}^{0}$-continuum 
contribution were made for 
$\Lambda_{D^{*0}} = 3.3$ and $3.5~{\rm GeV}$ in 
(\ref{DstarDgam_ff_exp}), which correspond to the lower and upper 
lines in the bands, respectively.
We used the BT potential for the excited $2P$ states, 
$\chi_{c0}(2P)$ and $\chi_{c2}(2P)$.
(From figure~5 of \cite{Babiarz:2025sld}).}
\label{fig:M34_BaBar_eff}
\end{figure}

Finally, in figure~\ref{fig:M34_BaBar_eff} we show our complete model 
results for $e^{+} e^{-} \to e^{+} e^{-} D \bar{D}$
including $D^{0} \bar{D}^{0}$ continuum,
$D^{+} D^{-}$ continuum, $\chi_{c0}(3860)$, and $\chi_{c2}(3930)$ 
contributions together with the efficiency-corrected (combined) 
BaBar data \cite{BaBar:2010jfn}.
In the calculation we used the Buchm\"uller-Tye potential
for the excited $2P$ states, $\chi_{c0,2}(2P)$.
The $D^{0} \bar{D}^{0}$-continuum contribution from 
$t/u$-channel vector-meson $D^{*}(2007)^{0}$ exchanges
depends on the parameter of the off-shell form-factor 
(\ref{DstarDgam_ff_exp})
[see equation~(2.39) of \cite{Babiarz:2025sld}].

\section{Conclusions}
\label{sec:Conclusions}

In reference~\cite{Babiarz:2025sld} we have concentrated on the possible observation
of excited $P$-wave charmonia 
in the $e^+ e^- \to e^+ e^- D \bar{D}$ reaction.
One of our goals was to estimate the cross-section of meson-pair 
continua for both $D^0 {\bar D}^0$ and $D^+ D^-$ channels separately.
We have considered $t$- and $u$-channel exchanges of the vector 
$D^{*0}$ and $D^{*\pm}$ mesons. The corresponding coupling constants
for the $D^* D \gamma$ couplings was obtained in \cite{Babiarz:2025sld} from 
the experimental radiative partial decay widths of $D^* \to D \gamma$.
For the $D^+ D^-$ channel, we also considered the continuum mechanism
via the $D^{\pm}$ $t/u$-channel exchanges plus contact term.
Comparing our model results with the experimental data 
of Belle \cite{Belle:2005rte} and BaBar \cite{BaBar:2010jfn}
we found a large contribution of the continuum to the 
$D^0 {\bar D}^0$ channel and a much smaller one for the 
$D^+ D^-$ channel.

Our analysis suggests that the broad bump at $M_{D \bar D} = 3.8$~GeV
observed in the $D^0 {\bar D}^0$ channel, and not observed 
in $D^+ D^-$, is rather due to the continuum mechanism.
We have shown that taking this resonance into account 
with $\Gamma_{\chi_{c0}} = 50$~MeV
improves the description of the near-threshold data,
in particular, the BaBar data 
for the $D^{+}D^{-}$ channel.
It is rather difficult to understand a large difference between
$\Gamma_{\chi_{c0}}$ and $\Gamma(\chi_{c0} \to D \bar{D})$
in the scenario with large $\Gamma_{\chi_{c0}}$ given in 
(\ref{4.0_pdg})
which leads to surprisingly small branching fraction
$B(\chi_{c0} \to D \bar{D})$.
This indicates that further theoretical and experimental research 
is needed to understand properties of the $\chi_{c0}(2P)$ state better.

We estimated the cross section for the $\chi_{c2}(3930)$
production in no tag conditions using $R_{2P}'(0)$ 
calculated within the light front quark model.
Calculations were performed for different 
potentials known from the literature.
We have obtained an integrated cross-section 
$\sigma(e^+ e^- \to e^+ e^- \chi_{c2}(3930)) \sim 1$~pb
using the Buchm\"uller-Tye (BT) potential.
From a comparison of our results 
for the $e^+ e^- \to e^+ e^- (\chi_{c2}(3930) \to D \bar{D})$ reaction
to the BaBar results, we found
the branching fraction $B(\chi_{c2}(3930) \to D \bar{D}) = 0.58 \pm 0.13$
(BT potential) and $0.38 \pm 0.08$ (Cornell potential).
The presented results strongly motivate a more detailed analysis of 
the $D \bar{D}$ production.
Hopefully, this will be possible with future Belle~II high-statistics data.

In reference \cite{Lebiedowicz:2026qff}, 
the interested reader may find predictions for the exclusive $D \bar{D}$ production, 
limiting to low invariant mass $M_{D \bar{D}} < 4.3$~GeV,
in ultraperipheral
ultrarelativistic heavy-ion collisions (UPCs)
at energies available at the LHC.
The nuclear cross sections for the $D^{0} \bar{D}^{0}$ and $D^{+} D^{-}$ channel
calculated at $\sqrt{s_{NN}} = 5.02$~TeV
and for $|\eta_{D}| < 2.5$,
$p_{t,D} > 0.2$~GeV, and $M_{D \bar{D}} < 4.3$~GeV,
are about 115~$\mu$b and 30~$\mu$b, respectively.
Therefore, the $D\bar{D}$ production in heavy-ion UPCs 
may provide new information compared 
to the presently available Belle and BaBar data.
It is also worth noting that the production
of open-charm meson pairs in UPCs 
is one of the potential opportunities to search 
for the lightest possible exotic charmonium state.

\end{document}